\documentclass[12pt]{article}
\usepackage{epsf}
\usepackage{graphics}

\begin{document}
\title
{A quantum gravitational model of redshifts}
\author{Michael A. Ivanov \\
Physics Dept.,\\
Belarus State University of Informatics and Radioelectronics, \\
6 P. Brovka Street,  BY 220027, Minsk, Belarus
\\ Email: ivanovma@gw.bsuir.unibel.by.}
\maketitle

\begin{abstract}
The main features of an alternative model of redshifts are
described here. The model is based on conjectures about an
existence of the graviton background with the Planckian spectrum
and a super-strong character of quantum gravitational interaction.
This model is connected with the assumed quantum mechanism of
gravity. A behavior of two theoretical functions of a redshift $z$
in this model - the geometrical distance $r(z)$ and the luminosity
distance $D_{L}(z)$ - and an existence of two different
cosmological horizons for any observer are discussed.
\end{abstract}

\section[1]{Introduction }
A commonly accepted hypothesis about the Dopplerian nature of
cosmological redshifts leads to a necessity to introduce dark
matter and dark energy into consideration to explain kinematically
the latest SNe 1a observations \cite{3}. As it was shown by the
author \cite{1,5}, a very specific apparent dimming of supernovae
may be interpreted in an essentially different way, without any
kinematics. This new approach is based on a few simple, but
partially unexpected, conjectures: there is the graviton
background with the Planckian spectrum having a small effective
temperature; gravitons are super-strong interacting particles; a
cross-section of interaction of a graviton with any particle is a
bilinear function of energies of both particles. It was also shown
\cite{2} that these conjectures may underlay a quantum mechanism
of classical gravity. It has a very interesting consequence: two
fundamental constants - the Hubble constant and the Newton
constant - {\it should be connected between themselves} in this
approach. The main features of this model of redshifts are
summarized in this short paper.
\section[2]{Redshifts as a quantum gravitational effect}
The isotropic graviton background is considered in the model to be
fulfilling a flat non-expanding universe. Gravitons should be
super-strong interacting ones to give a full magnitude of
cosmological redshifts. There are two effects due to collisions of
photons with gravitons. Because of forehead collisions, there
should exist a photon average energy loss that leads to a redshift
$z$ on a geometrical distance $r:$ $$z=\exp(ar)-1,$$ where
$a=H/c,$ and for the Hubble constant $H$ we have \cite{1}: $$H=
(1/2\pi) D \cdot \bar \epsilon \cdot (\sigma T^{4}),$$ $\bar
\epsilon$ is an average graviton energy, $\sigma$ is the
Stephan-Boltzmann constant, $T$ is an effective temperature of the
background, $D$ is a new constant. It is assumed here that  for
forehead collisions, a cross-section $\sigma (E,\omega)$ of
interaction of any particle with an energy $E$ and a graviton with
an energy $\omega$ is equal to:
$$\sigma (E,\omega)= D \cdot E \cdot \omega. $$
To have a reasonable value for the Hubble constant, it must be: $D
\sim 10^{-27} m^{2}/eV^{2}.$
\par
Another effect is caused by non-forehead collisions of photons
with gravitons, and leads to additional relaxation of any photonic
flux. This relaxation depends on the relaxation factor $b$ which
is equal to: $b=3/2 + 2/\pi = 2.137,$ as it is shown in details in
my recent paper \cite{7}. Both redshifts and this relaxation give
the following dependence of the luminosity distance $D_{L}$ on
$z:$
$$D_{L}=a^{-1} \ln(1+z)\cdot (1+z)^{(1+b)/2} \equiv
a^{-1}f_{1}(z).$$ This function fits well supernova observational
data by Riess et al. \cite{3} for small $z<0.5$ (see \cite{5}).
Discrepancies for higher $z$ would be understood as a result of
deformation of any spectrum it this redshift model due to a
non-zero value of an average graviton energy which is equal to
$\sim 10^{-3}$ eV by $T=2.7$ K. At present, there does not exist a
full theoretical description of this third effect due to the
graviton background.
\par
The Newton constant $G$ may be expressed in this approach via the
same quantities as $H$, and we have the connection between them
\cite{2}. From the latter, we get for $H:$ $H= 3.026 \cdot
10^{-18}s^{-1}=94.576 \ km \cdot s^{-1} \cdot Mpc^{-1}$ by $T=2.7
K.$
\par
\begin{figure}[th]
\epsfxsize=12.98cm \centerline{\epsfbox{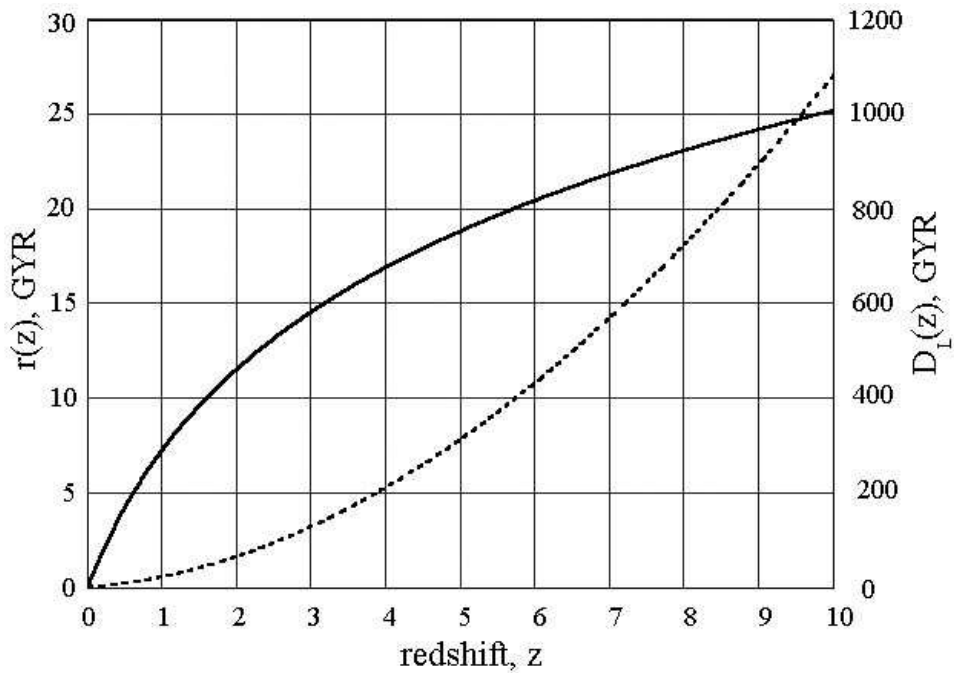}} \caption{The
geometrical distance, $r(z),$ (solid line) and the luminosity
distance, $D_{L}(z),$ (dashed line) - both in light GYRs - in this
model as functions of a redshift, z. The following theoretical
value for $H$ is accepted: $H= 3.026 \cdot 10^{-18}s^{-1}$.}
\end{figure}
By this value of $H$ (then a natural light unit of distances is
equal to $1/H \simeq 10.5$ light GYR), plots of two theoretical
functions of $z$ in this model - the geometrical distance $r(z)$
and the luminosity distance $D_{L}(z)$ - are shown on Fig. 1. As
one can see, for objects with $z \sim 10$, which are observable
now, we should anticipate geometrical distances of the order $\sim
25$ light GYR and luminosity distances of the order $\sim 1100$
light GYR in a frame of this model.
\par
\begin{figure}[th]
\epsfxsize=12.98cm \centerline{\epsfbox{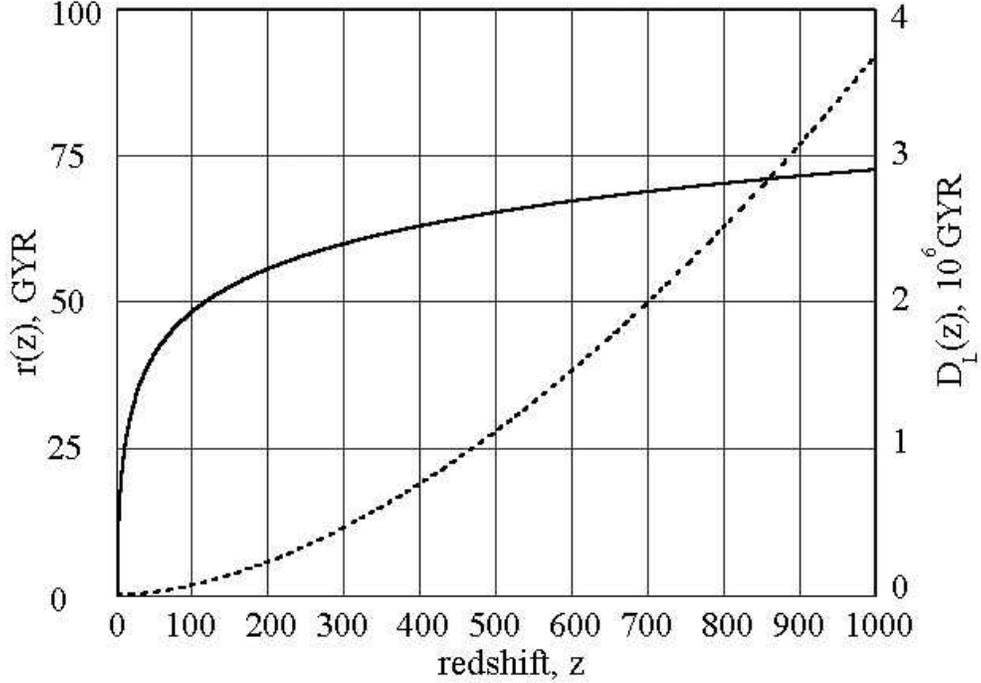}} \caption{The
same functions as on Fig. 1 (all notations are reserved), but for
the huge range of $z$.}
\end{figure}
We can assume that the graviton background and the cosmic
microwave one are in a state of thermodynamical equilibrium, and
have the same temperatures. CMB itself may arise as a result of
cooling any light radiation up to reaching this equilibrium. Then
it needs $z \sim 1000$ to get through the very edge of our cosmic
"ecumene" (see Fig. 2). Some other possible cosmological
consequences of existence of the graviton background were
described in \cite{8,2}.

\section[1]{Conclusion}
It is necessary to develop a theory of redshifts in this approach
taking into account the quantum nature of red-shifting process and
a non-zero value of an average graviton energy. Of course, a
verification of redshift's origin on the Earth, which would be
carried out with coming ultrastable lasers, will be of great
importance. In a case of the waited Dopplerian nature of
redshifts, one will get a negative result trying to detect a laser
beam frequency shift after a delay line. If the considered
conjecture about the gravitational origin of redshifts is true, a
result will be positive.
\par
It is interesting that in a frame of this model, every observer
has two own spheres of observability in the universe. One of them
is defined by maximum existing temperatures of remote sources - by
big enough distances, all of them will be masked with the CMB
radiation. Another, and much smaller, sphere depends on their
maximum luminosities - the luminosity distance increases with a
redshift much quickly than the geometrical one. An outer part of
the universe will drown in a darkness.

\end{document}